# Optical bistability in Er-Yb co-doped phosphate glass microspheres at room temperature

**Jonathan M. Ward[(a)]**
*Department of Applied Physics and Instrumentation, Cork Institute of Technology, Bishopstown, Cork, Ireland*
*Photonics Centre, Tyndall National Institute, Prospect Row, Cork, Ireland*

**Danny G. O'Shea**
*Physics Department, University College Cork, Ireland*
*Photonics Centre, Tyndall National Institute, Prospect Row, Cork, Ireland*

**Brian J. Shortt**
*Department of Applied Physics and Instrumentation, Cork Institute of Technology, Bishopstown, Cork, Ireland*
*Photonics Centre, Tyndall National Institute, Prospect Row, Cork, Ireland*

**Síle Nic Chormaic**
*Physics Department, University College Cork, Ireland*
*Photonics Centre, Tyndall National Institute, Prospect Row, Cork, Ireland*

We experimentally demonstrate optical bistability in $Er^{3+}$-$Yb^{3+}$ phosphate glass microspheres at 295 K. Bistability is associated with both $Er^{3+}$ fluorescence and lasing behavior, and chromatic switching. The chromatic switching results from an intrinsic mechanism exploiting the thermal coupling of closely-spaced energy levels, and occurs simultaneously with the intensity switching. A contrast ratio of 3.2 has been obtained for chromatic switching, and the intensity switching shows ratios of 2.4 for 550 nm and, 1.8 for the 660 nm fluorescence emissions, and 11 for the IR lasing at 1.5 $\mu$m. Concurrent with these observations, we investigate a temperature dependent absorption of pump power which exhibits bistable behavior. The influences of the host matrix on lasing and fluorescence mechanisms are highlighted.

---

[a] Jonathan.ward@cit.ie

**I. INTRODUCTION**

The first demonstration of optical bistability (OB) was reported by Gibbs in 1976[1] in a Na vapor and, since then, numerous other materials exhibiting the phenomenon have been studied, including $Yb^{3+}$ doped glasses and crystals, and semiconductors.[2-8] The mechanisms responsible for nonlinearity in glasses and crystals are varied, with many requiring cryogenic temperatures (typically < 40 K) to maintain the necessary low atomic decay rates in $Yb^{3+}$ dimer and monomer systems.[3] Alternative and more easily achievable mechanisms include photon avalanche and thermal avalanche,[2,5] suggesting that a wider range of materials can exhibit OB at temperatures approaching room temperature and above. However, to the best of our knowledge, no evidence of OB through these mechanisms in this temperature range has yet been observed. Bistable sensitized luminescence in Er-Yb:$CsCdBr_3$ was first shown by Redmond,[3] and subsequently by Ródenas in a Nd-Yb co-doped crystal.[4] They also described *chromatic* switching in addition to the more traditional *intensity* switching, though both mechanisms required temperatures below room temperature. To date, OB has also been predicted and observed in $Yb^{3+}$ doped oxide crystals, Cr-doped $LiSrGaF_6$ and $LiSrAlF_6$ crystals,[5] $Sm^{3+}$ doped glass microspheres,[6] and $Tm^{3+}$-$Yb^{3+}$ co-doped glass. The later system exhibited multiple hysteresis loops in the fluorescence intensity at room temperature.[7]

The interest in studying micron-sized spherical cavities stems from the potential they offer as novel components for experiments ranging from the very applied realization of all-optical networks to fundamental quantum optics experiments, whereby the microspheres can be used as ultra-high Q cavities for measurements based on the principles of cavity quantum electrodynamics.[9] Such microcavities can be either active or passive depending on the material used. For example, a sphere doped with the triply-ionized rare earth ions, $Er^{3+}$, can yield fluorescence emissions ranging from UV to IR through various upconversion mechanisms [10,11] due to the close proximity of the numerous energy levels in the ion. On the other hand we have previously shown[12] that certain emissions from an erbium-doped, fluoride glass microcavity are coincident with the resonance wavelengths of the D-2 transitions in caesium and rubidium - atoms that are routinely used in atom optics experiments. Microcavities that exhibit optical bistability are interesting for generating optical switches for all-optical computing, and miniature C-band laser sources are important for telecommunications applications.[13] These factors have resulted in significant research focusing on the characterization of microcavities in recent years.

The concept of an $Er^{3+}$-$Yb^{3+}$ co-doped glass laser was demonstrated[14] in 1965 as a means of optimizing the $Er^{3+}$ emission cross-section while simultaneously ensuring an optimum $Yb^{3+}$ absorption cross-section. This overlap alleviates the difficulty of trying to directly pump the narrow $Er^{3+}$ absorption band.[15] Phosphate glass has been investigated as a host matrix for rare-earth ions due to its favorable properties, such as: (i) the possibility of obtaining large dopant concentrations (up to $1.8 \times 10^{21}$ ions/cm$^3$ for $Yb^{3+}$, and ~$10^{19}$ ions/cm$^3$ for $Er^{3+}$) compared to silicate, borate and fluoride glasses, (ii) its large absorption band in the near-infrared region, (iii) its large emission cross-section at 1.5 $\mu$m, and (iv) the low back energy transfer from $Er^{3+}$ ions to $Yb^{3+}$ ions.[16,17,18] The spectral characteristics are also especially beneficial for C-band lasing; the intermediate lasing level, $^4I_{11/2}$, has a high non-radiative relaxation rate (< 1 $\mu$s lifetime, maximum phonon energy of 1300 cm$^{-1}$) to the $^4I_{13/2}$ level compared with silica (maximum phonon energy of 1190 cm$^{-1}$), and the long lifetime of the $^4I_{13/2}$ level of about 8.45 ms facilitates population inversion and high gain.[13] The larger phonon energy of phosphate glass has a negative effect on the upconversion efficiency compared to fluoride glass (with a maximum phonon energy of 600 cm$^{-1}$). This, however, is counter-balanced by the large dopant concentrations and wide absorption cross-section of the $Yb^{3+}$ sensitizer. In addition, phosphate glass has better optomechanical properties compared to other glasses used for OB, such as CsCdBr$_3$.

Numerous papers have reported the linear excitation power dependence of green and red upconversion processes in bulk Er$^{3+}$-Yb$^{3+}$ co-doped phosphate glass, waveguides and - to a lesser extent - microspheres.[19,20] Here we report the first observation of chromatic and intensity OB in phosphate glass (Schott IOG-2) at room temperature. To our knowledge, simultaneous intensity OB for multiple emissions and chromatic OB has not previously been reported for a single material. We present experimental results on 1.56 $\mu$m lasing and propose suitable upconversion mechanisms for the three color emission. For clarity, our results are organized in two sections: fluorescence and lasing emissions are discussed in section III, and the results of the OB measurements are discussed in section IV.

## II. EXPERIMENT

The IOG-2 glass used is doped with 2wt% Er$_2$O$_3$ (1.7×10$^{20}$ ions/cm$^3$) and co-doped with 3wt% Yb$_2$O$_3$ (2.5×10$^{20}$ ions/cm$^3$). We study lasing and upconversion fluorescent emissions following CW pumping with a tunable 980 nm laser-diode (spectral width ~ 1 nm). IOG-2 glass, with a low glass transition temperature of around 700 K,[16] is ideal for producing microspheres with diameters of between 30-70 $\mu$m using a microwave plasma torch.[20] A detailed description of the experimental approach has been published elsewhere.[21] Efficient coupling of the pump into the microsphere is attained by using adiabatically tapered fibers fabricated using a direct heating technique.[22] We use 3 $\mu$m diameter, 980 nm SMF-28 in the experiment for studying the upconversion and lasing processes, and we use 1 $\mu$m diameter, 1550 nm SMF-28 in the OB experiment. The transmission loss in the tapered SMF-28 fiber is typically less than 0.1 dB/cm. The alignment of taper and microsphere is optimized by adjusting the relative positions of the two, whilst maximizing the 1550 nm emissions. These lasing emissions are monitored by connecting one end of the pump fiber taper to an optical spectrum analyzer. During alignment we also monitor the transmission through the pump fiber and typically 10–15% of the pump light is coupled into the microsphere with the 1550 nm fiber taper and only about 1% with the 980 nm fiber taper. All upconversion fluorescence spectra were acquired by free space coupling into an Ocean Optics 2000 spectrometer.

## III. Lasing and Fluorescence in $Er^{3+}$-$Yb^{3+}$

Figure 1 shows the presence of switching in the fluorescence spectrum of a 35 $\mu$m diameter sphere before and after the switching positions, and the inset shows efficient fluorescent upconversions ranging from violet to red wavelengths for a pump power of 8 mW. We note four distinct emission bands corresponding to erbium transitions at 405 nm ($^2H_{9/2} \to {}^4I_{15/2}$), 520 nm ($^2H_{11/2} \to {}^4I_{15/2}$), 550 nm ($^4S_{3/2} \to {}^4I_{15/2}$), and 660 nm ($^4F_{9/2} \to {}^4I_{15/2}$). The red emission is stronger than red emissions produced in other singly doped glasses investigated in our laboratory, such as Er:ZBLALiP and Er:ZBNA, due to the much larger $Yb^{3+}$ absorption cross-section compared to $Er^{3+}$. One would expect the slope of the fluorescence to exhibit a simple power law dependence, $I_{emission} \propto I_{excitation}^a$, where $a$ is the number of pump photons required to produce each emitted photon, thus reflecting the multiphoton nature of the upconversion process. However, this law fails in the presence of OB, as will be explained in section IV.

    Analysis of the whispering gallery modes in the IR fluorescence in the inset of Fig. 2 enables us to measure the free spectral range, $\nu_{FSR}$, of the sphere modes to be 1.32 THz ($\equiv$ 10.4 nm). This agrees closely with the calculated value of $\nu_{FSR} = c/\pi ND \approx 1.27$ THz ($\equiv$ 10.2 nm), indicating a microsphere diameter, $D \approx 48$ $\mu$m. This has been confirmed with an optical microscope measurement of 50 ± 2 $\mu$m. The refractive index of IOG-2 at room temperature is 1.508 at 1540 nm. Eccentricity, $\varepsilon$, in the microsphere results in splitting of the azimuthal modes,

$\Delta\omega_{ecc}$, by about 15 GHz corresponding to an eccentricity of around 2.7% for $l = |m|$, calculated from $\varepsilon = (\Delta\omega_{ecc}/\omega_{nml}) \cdot (l^2/|m| - 1/2)$, where $l$ and $m$ are the angular mode numbers, and $\omega_{nml}$ is the angular frequency of the mode. The measured spacing between TE and TM modes of 9.0 nm around 1550 nm agrees well with the calculated value of 8.5 nm. This microsphere has a peak lasing emission of 44 $\mu$W for a pump power of ~ 10 mW coupled into the fiber taper and, in general, we record lasing thresholds of less than 1 mW.

Energy transfer from the $^4F_{5/2}$ level in $Yb^{3+}$ to the $^4I_{11/2}$ level in $Er^{3+}$ is only about 80% efficient, resulting in the remaining 20% being dissipated in the host matrix as heat.[17] The $^2H_{11/2}$ level and the $^4S_{3/2}$ level can be considered to be in quasi-thermal equilibrium since $k_B T \approx 200$ cm$^{-1}$ is comparable to the maximum phonon energy of 1300 cm$^{-1}$ for phosphate glass.[19] As such, only one phonon is required to bridge the energy difference between the two green levels, thereby populating the $^2H_{11/2}$ level at room temperature. The thermalization of the $^2H_{11/2}$ level by the $^4S_{3/2}$ level has a temperature dependent effect on the ratio of the radiative emissions from these two levels as well as the excited state lifetimes.[23,24] The multi-phonon emission rate at a temperature, $T$, is

$$K_{mp}(T) = K_{mp}(0)\left[1 - e^{-\hbar\omega/k_B T}\right]^{-p} \quad (1)$$

where $k_B$ is Boltzmann's constant, $\omega$ is the phonon frequency, and $p$ is the number of phonons required to bridge the energy gap, $\Delta E$, between the levels and is given by $\Delta E/\hbar\omega$, where $\hbar\omega$ is the phonon energy.[24] For phosphate glass, this emission rate is $K_{mp}(295\ K) \approx 10^{11}$ s$^{-1}$, which is significantly higher than the value of $10^8$ s$^{-1}$ reported for silica.[25] The energy difference between the $^4S_{3/2}$ level and the next lowest level, $^4F_{9/2}$, requires three phonons, therefore making this level far less likely to be populated from the $^4S_{3/2}$ level.

The emissions detected from the IOG-2 microsphere are due to upconversion processes involving multiple pump laser photons and/or lattice phonons. Two upconversion mechanisms which must be considered when trying to understand the origin of the fluorescence results are excited state absorption (ESA) and energy transfer upconversion (ETU).[26,27] These are shown in the energy level diagram in Fig. 3. The combination of reasonably high cavity quality factor (typically 2-5 × $10^4$, measured as the FWHM of an individual 1.5 $\mu$m lasing peak and limited by the resolution of the optical spectrum analyzer) and strongly localized electromagnetic field in the form of a whispering gallery mode (mode volume ~ 3000 $\mu m^3$) serves to significantly enhance the probability that an excited ion will absorb further pump photons. As mentioned, there is a resonant energy transfer (ET) from the $Yb^{3+}$ sensitizer ($^2F_{5/2} \rightarrow {}^2F_{7/2}$) to the $Er^{3+}$ ion ($^4I_{15/2} \rightarrow {}^4I_{11/2}$) followed by ESA from the $^4I_{11/2}$ level to the $^4F_{7/2}$ level. However, rapid, non-radiative relaxation to the $^4I_{13/2}$ level is also possible. In spite of a large energy mismatch of about 1450 $cm^{-1}$ for the $^4I_{13/2} \rightarrow {}^4F_{9/2}$ transition, the long lifetime of the $^4I_{13/2}$ state and associated large population ensures that the ETU mechanism is adequately efficient.[28,29] Due to the close spacing of the $^4F_{7/2}$ and $^4S_{3/2}$ levels, the population of $^4F_{7/2}$ readily decays non-radiatively to the $^4S_{3/2}$ level, whereby the thermal mechanism described previously populates the $^2H_{11/2}$ level. Finally, a second ESA from the $^4S_{3/2}$ level up to the $^2G_{7/2}$ level is followed by non-radiative relaxation down to the $^2H_{9/2}$ level and the subsequent radiative decay to the ground state generates a photon at 410 nm (violet).

At the high concentrations of $Er^{3+}$ and $Yb^{3+}$ in this work the inter-ion separation reaches a critically small radius of ~ 3 nm for $Er^{3+}$ ions and ~ 2 nm for $Yb^{3+}$ ions. This is close to the value of 2.12 nm for $Yb^{3+}$-$Yb^{3+}$ ET's and an estimated 1.5-2.0 nm for $Yb^{3+}$-$Er^{3+}$ ET's as determined from Förster-Dexter theory,[30,31] dramatically enhancing the probability of $Yb^{3+}$-$Yb^{3+}$,

Yb$^{3+}$-Er$^{3+}$ and, presumably, Er$^{3+}$-Er$^{3+}$ energy transfers. The critical radius, $R_{sx}$, is determined by the overlap of the emission and absorption cross-sections and is given by

$$R_{sx}^6 = \frac{3c\tau_s}{8\pi^4 n^2} \int \sigma_{ems}^s(\lambda)\sigma_{abs}^x(\lambda) d\lambda \qquad (2)$$

where $c$ is the velocity of the photons, $\tau_s$ = 1.4 ms is the fluorescence decay time of the unperturbed sensitizer, $n$ is the refractive index, $\sigma_{ems}$ is the emission cross-section and $\sigma_{abs}$ is the absorption cross-section in Fig. 4, as calculated from absorbance measurements with a bulk sample of IOG-2. The script $s$ stands for the sensitizer, i.e. the Yb$^{3+}$ ions, and the script $x$ stands for either the sensitizer or the Er$^{3+}$ acceptor ions. For non-radiative dipole-dipole interactions the energy transfer rate rapidly increases according to the inverse of the ion separation to the sixth power.

The strength of the red emission cannot be wholly explained by multi-phonon relaxation from the $^4S_{3/2}$ level to the $^4F_{9/2}$ level, due to the low relaxation rate. Notwithstanding the ESA process already mentioned, two ETU cross-relaxation channels can explain the strength of the red emission relative to the green emissions. The first ETU is via $^4I_{11/2}(B)+{}^4I_{13/2}(A) \rightarrow {}^4I_{15/2}(B)+{}^4F_{9/2}(A)$ and the second channel is via $^2H_{11/2}(B)+{}^4I_{11/2}(A) \rightarrow {}^4F_{9/2}(B)+{}^4F_{9/2}(A)$ where $A$ and $B$ denote the two erbium ions involved (Fig. 3). Therefore, ions are removed from the green levels and transferred to the red $^4F_{9/2}$ level. The two possible cross-relaxation processes feeding the $^4F_{9/2}$ level deplete the intermediate $^4I_{11/2}$ level and the metastable $^4I_{13/2}$ level, thereby placing the 1.5 $\mu$m emission in competition with the red emission and the other upconversion processes. We find that the power of the red emission is, typically, twice as high in microspheres that exhibit no lasing (due to excessive inhomogeneities in the cavity) compared to those that exhibit lasing.

## IV. Optical Bistability

## A. Switching Contrast

A very promising feature of these microcavity resonators is the high contrast switching observed in the fluorescence emissions and, particularly, in the IR lasing; the ratio of the emissions before and after switching (c.f. Fig. 5) is 2.4 for the green, 1.8 for the red, and 11 for the IR emissions. The violet emission is too weak to be considered in these measurements. As the pump power is increased, the emissions intensities shown in Fig. 5 remained almost constant until some critical power was reached, beyond which a sudden and dramatic rise in emission intensity was observed. The critical power depends on the taper-sphere coupling efficiency. Increasing the pump power even further caused the emission to level out once more. A reduction in pump power clearly demonstrates a wide bistable region, where the emission can have two intensity values depending on the history of the input power. It should be noted that each graph in Fig. 5 is taken for a different sphere and the switching occurs at a different input power for the three spheres.

## B. Temperature Measurements

As alluded to earlier, the strong temperature dependence in IOG-2 is due to the inefficiency of the energy transfer from the $Yb^{3+}$ sensitizer to the $Er^{3+}$ acceptor. A sudden rapid rise in the 520 nm ($^2H_{11/2}$) emission intensity as compared to the 550 nm ($^4S_{3/2}$) intensity was observed when the pump power was increased above the upper switching position as shown in Fig. 1. As the pump power was reduced below the lower switching position, there was a sudden decrease in intensity ratio. Boltzmann statistics can simply describe the strong thermal coupling between these levels, and can lead to this type of switching. The intensity ratio is a function of the internal cavity temperature and is described as

$$\frac{I(^2H_{11/2})}{I(^4S_{3/2})} = \frac{r_H g_H \hbar \omega_H}{r_S g_S \hbar \omega_S} \exp\left(\frac{\Delta E}{k_b T}\right), \qquad (3)$$

where *I* is the integrated emission intensity for a particular level, *r* is the total spontaneous emission rate, *g* is the *(2J + 1)* multiplicity (or degeneracy) of each manifold, $\hbar\omega_H$ ($\hbar\omega_S$) is the energy of level $^2H_{11/2}$ ($^4S_{3/2}$), and *ΔE* is the energy separation between the levels.[24,25] According to eq. (3), as the temperature is increased, the $^2H_{11/2}$ level is more efficiently populated and an increasingly larger fraction of the $^4S_{3/2}$ population is rapidly promoted to the upper level and, consequently, the ratio of the emissions from the two states inverts so that the $^2H_{11/2}$ emission becomes stronger. This chromatic switching can yield intensity ratios of up 3.2 for the 520 nm ($^2H_{11/2}$) and 550 nm ($^4S_{3/2}$) emissions. According to eq. (3) this indicates a sphere temperature just below the glass transition value of 700 K. This temperature seems unreasonably high and there was no evidence of thermal stresses, fracturing of the glass or other defects in the microsphere during these measurements.

## C. External Heating of the Microsphere

The temperature dependence of the OB was examined by placing a platinum heater and thermocouple close to the microsphere and heating it to around 345 K. The pump power transmitted past the taper-sphere junction was recorded at the same time as the microsphere emissions, while cycling the pump power from low to high and back again. The nonlinear absorption of the pump photons (c.f. Fig. 6) is synchronized with switching in the lasing and fluorescence emissions. Comparison of the two plots taken at room temperature and at 345 K shows that an increase in external temperature causes the upper knee at 17 mW input power to shift down to 13 mW, while the lower knee at 7 mW only shifts slightly to the right, thereby shortening the bistable region. The pump power was cycled at a slow rate of ~ 1 mW/min due to the thermal response time of the microsphere.

The results presented here clearly show that IOG-2 microspheres exhibit bistable switching behavior under vastly different conditions to those reported for silica and semiconductor cavities, whereby the Kerr effect or thermal broadening/narrowing of the resonance line, together with high Q-factors ($Q>10^8$) and a tunable, narrow linewidth laser, provided the nonlinearity.[32] Two significant differences are found in our experiments: we use microspheres with loaded cavity Q's typically in the range $2–5 \times 10^4$, and the pump laser has a linewidth of 1 nm and is not locked to individual cavity resonances. Another possible nonlinearity mechanism - the thermo-optic effect - was used to explain OB in a Si resonator;[8] however, this effect is twenty times smaller for the IOG-2 glass presented in this paper.

We propose that the OB effects observed here can be explained by the thermal avalanche theory of Gamelin *et al.*[2] This theory relies on nonlinear absorption in the $Yb^{3+}$ ions with increasing temperature and predicts bistable power absorbance in the microsphere as a function of pump power in the taper as shown in Fig. 6. Measurements of the dependence of the $Yb^{3+}$ absorption coefficient would require it to be a nonlinear function of internal microsphere temperature above 295 K. As phonons are released into the lattice from $Yb^{3+}$ excited and ground states, the temperature increases, thereby increasing absorbance and leading to a further increase in temperature. This cyclic process causes a thermal avalanche for high enough pump power. Another possible explanation may be a nonlinear upconversion rate either for $Yb^{3+}$-$Yb^{3+}$, $Yb^{3+}$-$Er^{3+}$ or $Er^{3+}$-$Er^{3+}$ energy transfers.[33] This has been observed in high concentration $Er^{3+}$ doped fibers, where the nonlinearity is dependent on population inversion of the lasing levels, and the signal and emission rates. Temperature dependent effects in the upconversion fluorescence in Er-Yb $Ga_2S_3$:$La_2O_3$ chalcogenide glass and germanosilicate optical fibers have been attributed to an exponential increase in the $Yb^{3+}$ absorption cross-section elsewhere.[34] Apart from these observations, Guillot-Nöel *et al.*[35] have described bistability in the regime of strong $Yb^{3+}$-$Yb^{3+}$ coupling, in which case cooperative luminescence would be expected. Unfortunately, we have not been able to discount this effect. However, if it is present, the luminescence would likely be very weak compared to the single ion $Er^{3+}$ transitions. It is reasonable to assume the back transfer rate is negligible due to the high phonon energy of the host lattice, and is, therefore, unlikely to be the source of nonlinearity.

**V. CONCLUSIONS**

In conclusion, we have demonstrated a multi-wavelength, upconversion, microsphere light source with optical bistability and identified the $Er^{3+}$ transitions and fluorescence mechanisms involved. Factors affecting the dynamics of visible fluorescence and C-band lasing emission have been examined and show that the thermal properties of IOG-2 glass play an important role in the microsphere performance. Our calculations show that the close proximity of the ions in our glass greatly enhances the probability of sensitizer-sensitizer and sensitizer-acceptor energy transfers.

The first observation of chromatic and intensity OB in phosphate glass (Schott IOG-2) at room temperature is reported. These results show that there are two possible temperatures in the bistable region and it is also possible to have two emission intensities for the same excitation power. Chromatic bistability is dependent on the presence of intensity bistability due to the energy coupling between the green emitting levels, $^2H_{11/2}$ and $^4S_{3/2}$. Therefore the switching positions are the same for both types of bistability. The OB shows high contrast switching ratios. The intensity switching shows ratios of 2.4 for the green, 1.8 for the red fluorescence emissions, and 11 for the IR lasing, while the chromatic switching ratios ($^2H_{11/2}/^4S_{3/2}$) are as high as 3.2. We have found that the switching position is dependent on the microsphere temperature. The improved optomechanical properties of IOG-2 compared to other glasses used for OB, such as $CsCdBr_3$, makes this glass appealing for all-optical logic elements in optical engineering applications.

A microcavity resonator offers substantial miniaturization, greatly reduces the power required for switching, and allows for all emissions to be easily fiber coupled.[12,22] Consequently, such readily accessible optical switching features are important for applications in all-optical computing. Our observations may be tentatively explained in terms of a nonlinear, temperature dependent, absorption coefficient for the $Yb^{3+}$ sensitizer ions. We are currently modeling our data based on this assumption.

## VI. ACKNOWLEDGEMENTS

This work is funded by Science Foundation Ireland under grant 02/IN1/128. The authors acknowledge P. Féron and L. Ghisa from ENSSAT for providing the IOG-2 microspheres. D. O'Shea acknowledges support from the Irish Research Council for Science, Engineering and Technology through the Embark Initiative RS/2005/156.

Fig. 1 Upconversion fluorescence spectrum for an $Er^{3+}$-$Yb^{3+}$ co-doped IOG-2 microsphere. The main figure shows the spectrum before switching (dashed line) and after switching (solid line) obtained by pumping with a 1550 nm fiber taper. The ratio of emission from the $^2H_{11/2}$ to $^4S_{3/2}$ levels after switching is 3.2. The inset shows a spectrum obtained with a 980 nm fiber taper. The violet emission has been scaled up by a factor of 5 for clarity.

Fig. 2 Lasing spectrum for a $Er^{3+}$-$Yb^{3+}$ co-doped IOG-2 microsphere. The inset shows whispering gallery mode structure for a 50 $\mu$m diameter microsphere.

Fig. 3. Energy level diagram and fluorescence mechanisms with radiative (solid lines) and non-radiative transitions (wiggly lines). (1) Upconversion based on GSA and ESA in ion A only, (2) radiative and non-radiative decays, (3,4) energy transfer processes from ion A to ion B populating the $^4F_{9/2}$ level in ion A, (5) GSA and ESA in ion B. (CR) Cross-relaxation, (ET) Energy Transfer, (GSA) Ground State Absorption, (ESA) Excited State Absorption.

Fig. 4. Absorption and emission cross-sections for $Er^{3+}$-$Yb^{3+}$ co-doped IOG-2 bulk glass.

Fig. 5. Intensity bistability for green $^2H_{11/2}$ state, red $^4F_{9/2}$ state, and 1.5 $\mu$m lasing from the $^4I_{13/2}$ state taken for three different spheres. Label (1) in the green emission corresponds to a spectrum taken before switching and label (2) corresponds to the spectrum taken after switching such as those labeled in Fig. 1.

Fig. 6. Bistable absorption of pump power at 295 K and 345 K. The data at 345 K has been offset for clarity. As the sphere temperature was increased from 295 K to 345 K, the knee at 17 mW input power shifted to the left, while the lower knee at 7 mW only shifted slightly to the right, thereby shortening the bistable region. Label (1) indicates the region before switching and label (2) indicates the region after switching.

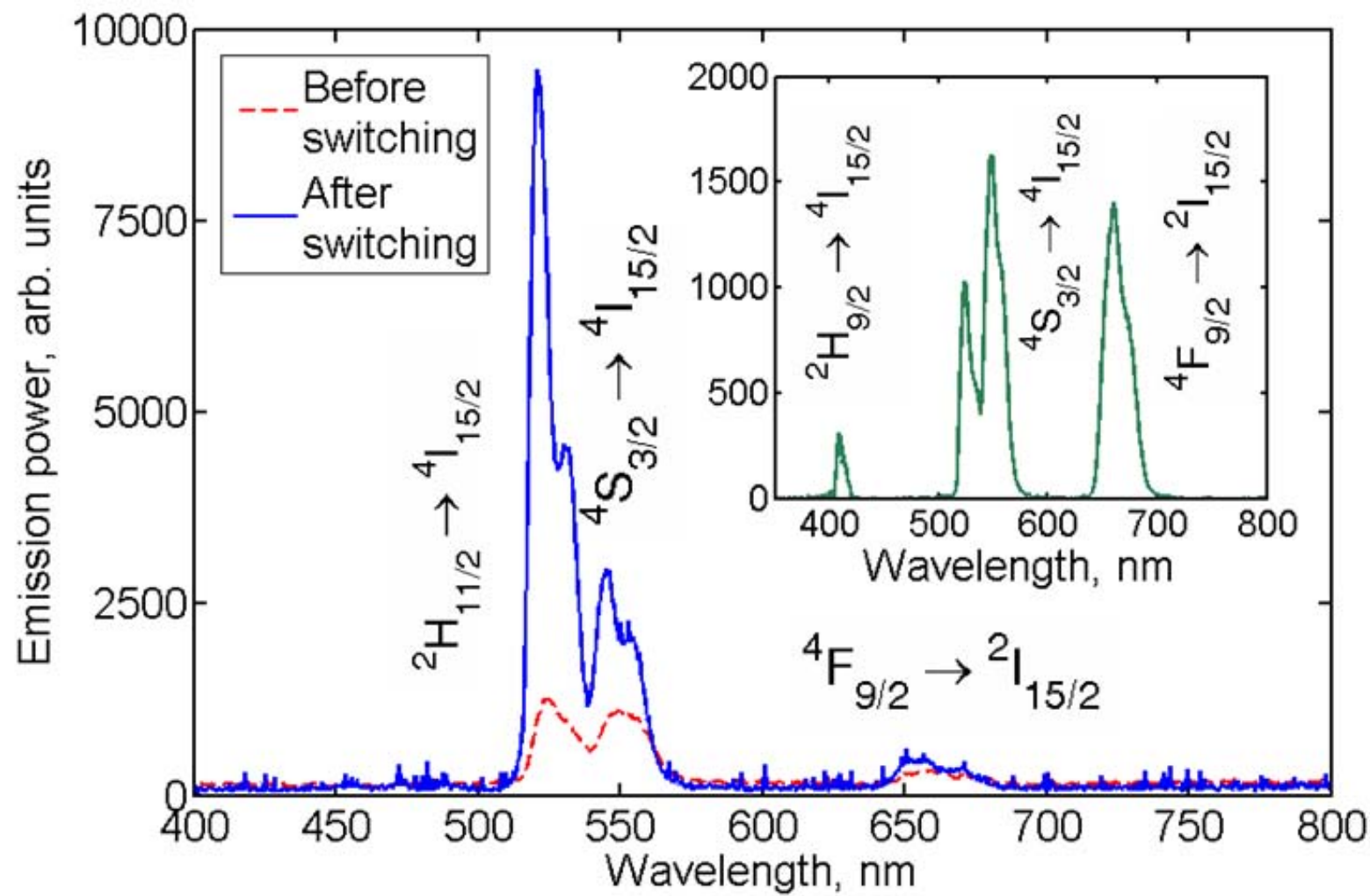

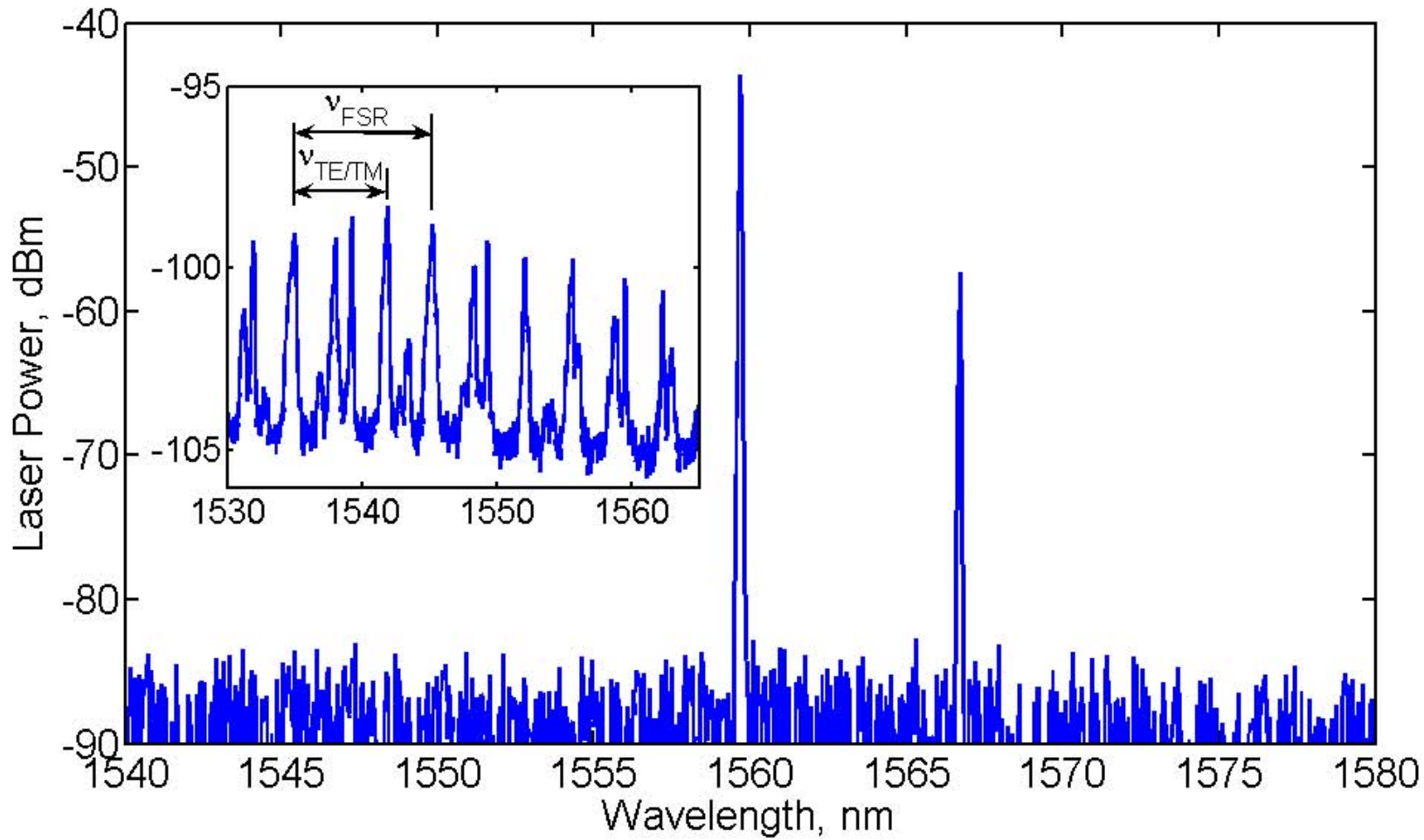

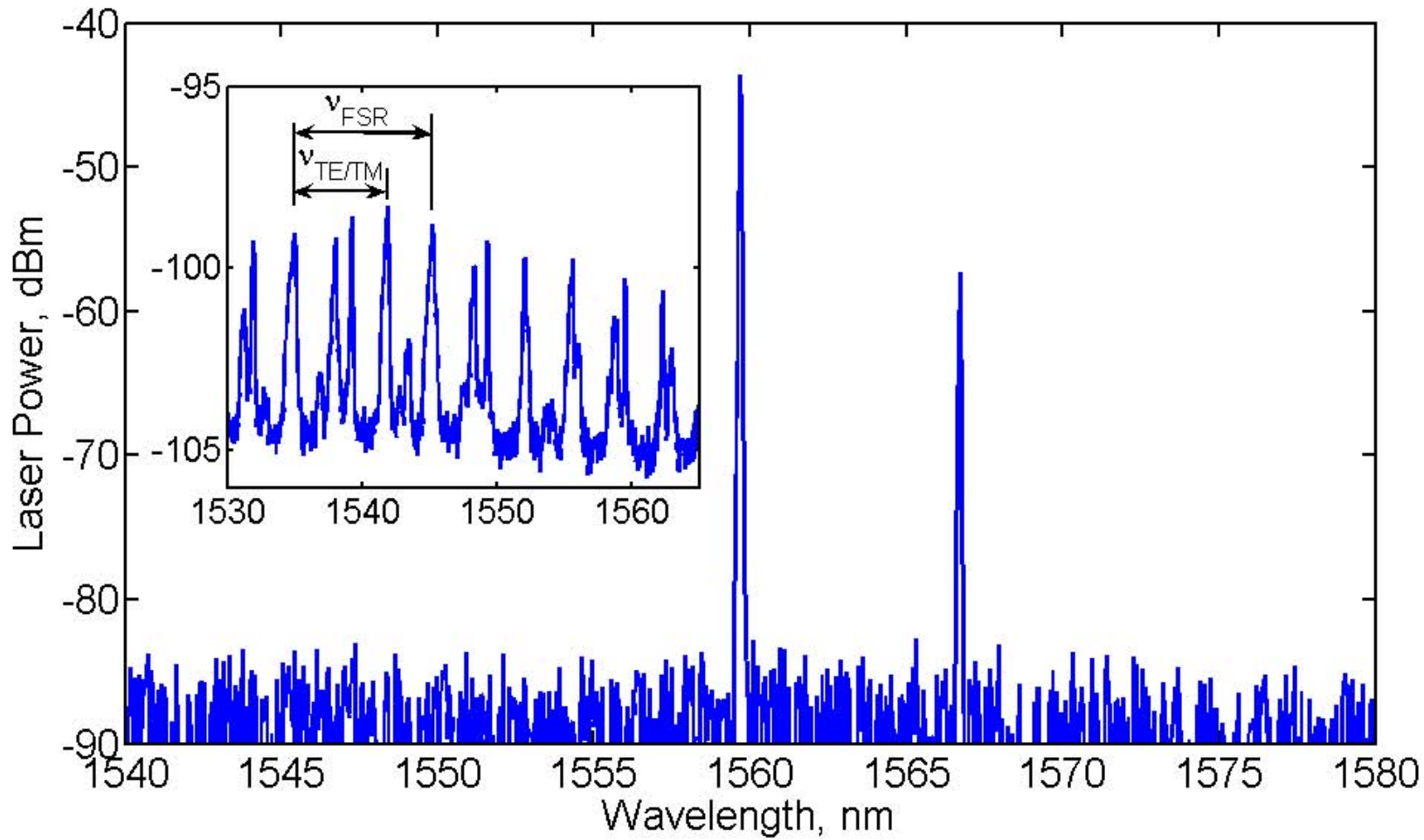

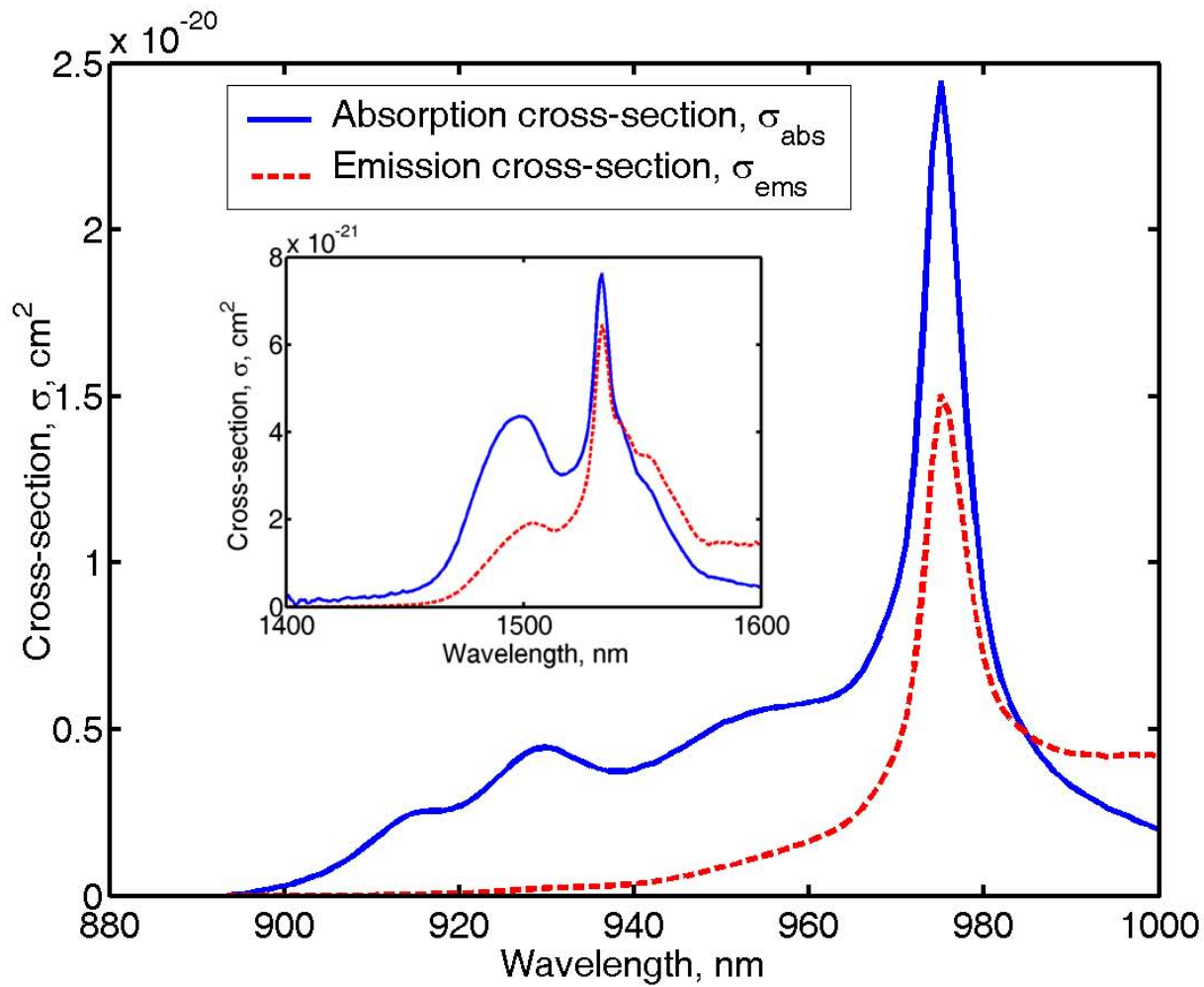

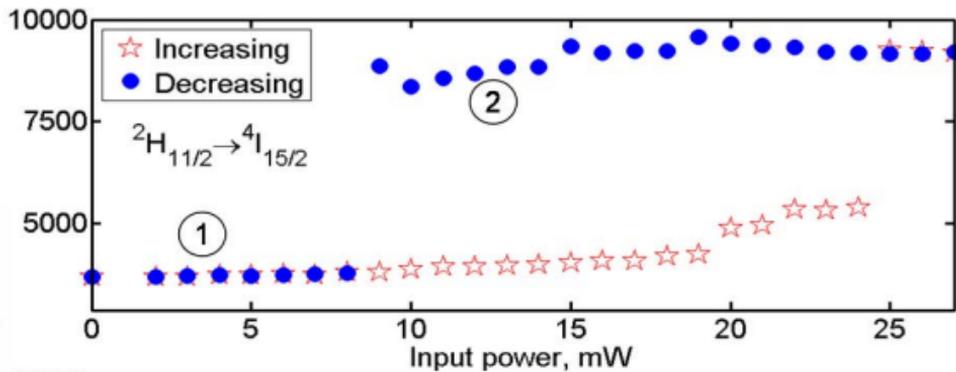
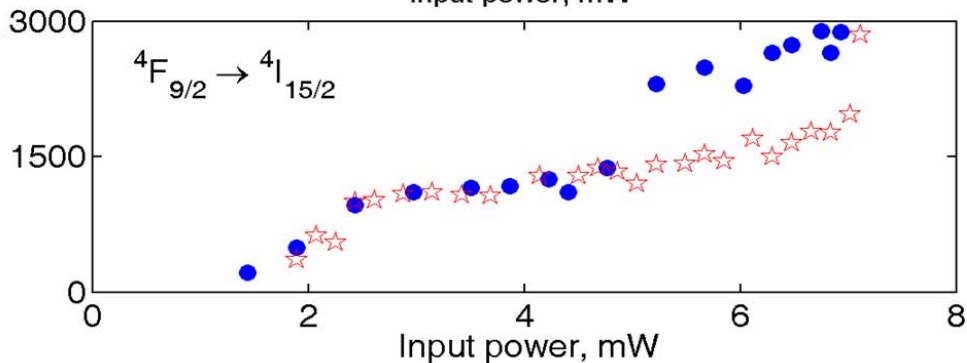
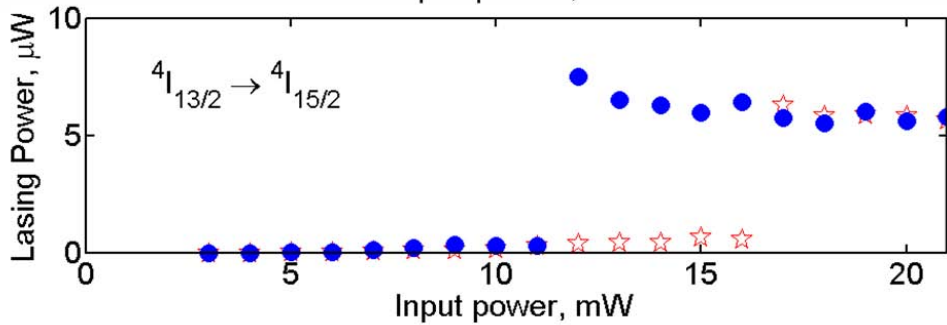

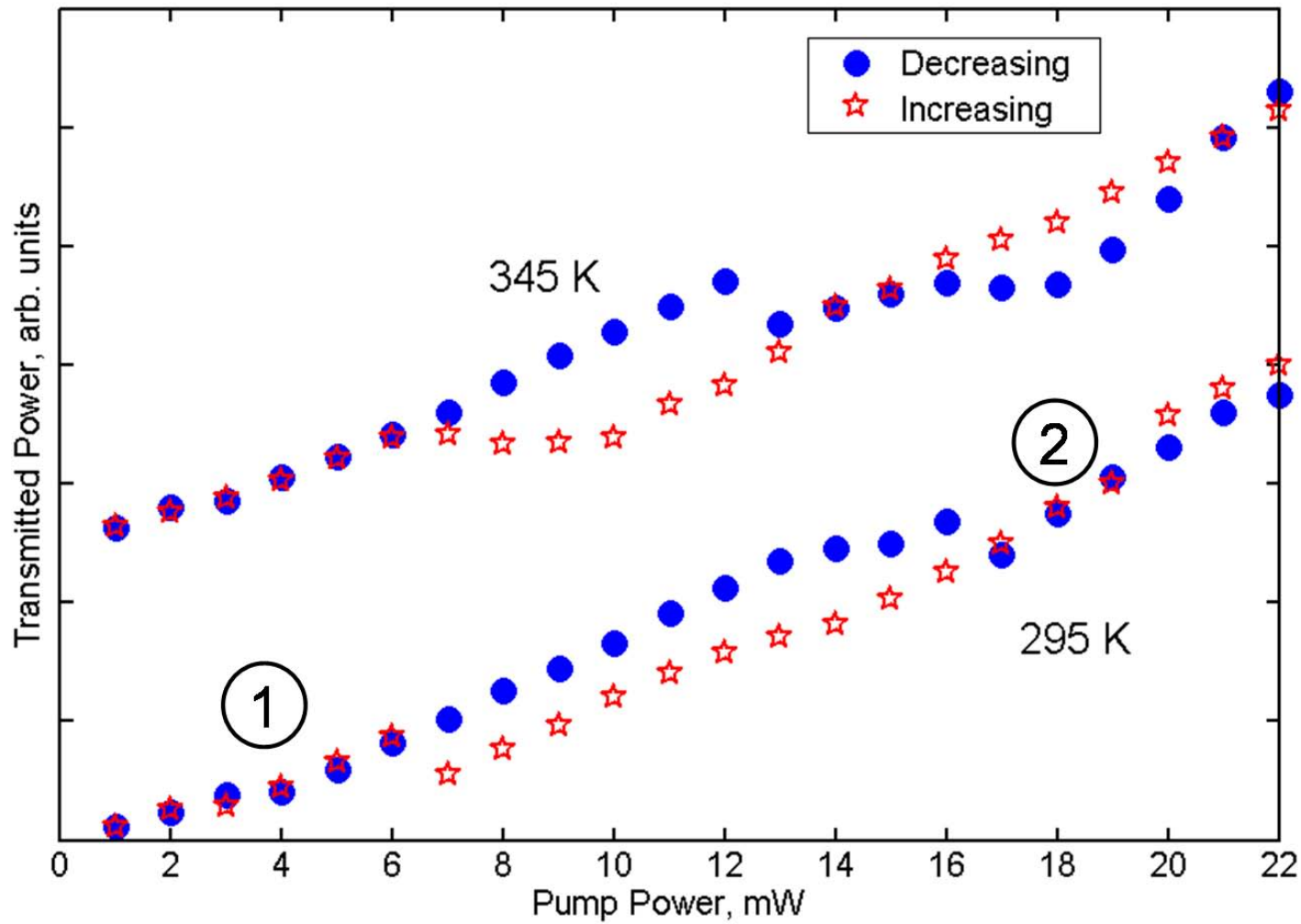